%
\catcode`@=11					
\let\notglobal=\relax
\let\notouter=\relax
\notouter\def\newcount{\alloc@0\count\countdef\insc@unt}
\notouter\def\newdimen{\alloc@1\dimen\dimendef\insc@unt}
\notouter\def\newskip{\alloc@2\skip\skipdef\insc@unt}
\notouter\def\newmuskip{\alloc@3\muskip\muskipdef\@cclvi}
\notouter\def\newbox{\alloc@4\box\chardef\insc@unt}
\notouter\def\newtoks{\alloc@5\toks\toksdef\@cclvi}
\notouter\def\newhelp#1#2{\newtoks#1#1\expandafter{\csname#2\endcsname}}
\notouter\def\newread{\alloc@6\read\chardef\sixt@@n}
\notouter\def\newwrite{\alloc@7\write\chardef\sixt@@n}
\notouter\def\newfam{\alloc@8\fam\chardef\sixt@@n}
\notouter\def\newinsert#1{\global\advance\insc@unt by\m@ne
  \ch@ck0\insc@unt\count
  \ch@ck1\insc@unt\dimen
  \ch@ck2\insc@unt\skip
  \ch@ck4\insc@unt\box
  \allocationnumber=\insc@unt
  \global\chardef#1=\allocationnumber
  \wlog{\string#1=\string\insert\the\allocationnumber}}
\notouter\def\newif#1{\count@\escapechar \escapechar\m@ne
  \expandafter\expandafter\expandafter
   \edef\@if#1{true}{\let\noexpand#1=\noexpand\iftrue}%
  \expandafter\expandafter\expandafter
   \edef\@if#1{false}{\let\noexpand#1=\noexpand\iffalse}%
  \@if#1{false}\escapechar\count@} 
\def\alloc@#1#2#3#4#5{\global\advance\count1#1 by 1
  \ch@ck#1#4#2%
  \allocationnumber=\count1#1
  \notglobal#3#5=\allocationnumber
  \wlog{\string#5=\string#2\the\allocationnumber}}
\catcode`@=12					


%
%

\newbox\IPtemp				
\newdimen\IPtempd			
\newdimen\IPspace\IPspace=.5em		
\newdimen\IPleft\IPleft=0pt		
\newif\ifIPtemps			
\newif\ifnotIPtemps			
%
%
%
\long\def\IP#1#2{{%
\setbox\IPtemp=\hbox{\noindent#1\hbox to \IPspace{}\hfil}%
%
\IPtempsfalse%
\notIPtempstrue%
\ifdim\IPleft>0pt%
	\ifdim\wd\IPtemp>\IPleft%
		\relax%
		\IPtempstrue%
		\notIPtempsfalse%
	\fi%
\fi%
\ifnotIPtemps\ifdim\wd\IPtemp<\IPleft%
	\setbox\IPtemp=\hbox{\hbox to \IPleft{\unhcopy\IPtemp}}%
\fi\fi%
%
\IPtempd=\the\hsize%
\ifnotIPtemps					
	\advance\IPtempd -\wd\IPtemp		
	\noindent
	\hbox{\unhbox\IPtemp\vtop{\hsize=\IPtempd\noindent#2}}%
\else						
	\advance\IPtempd -\IPleft%
	\noindent%
	\vbox{%
		\hbox{\unhbox\IPtemp}		
		\hbox{\hbox to \IPleft{}\vtop{\hsize=\IPtempd\noindent#2}}%
	}%
\fi%
\par%
}}
\font\eightrm=cmr8
\def\bold#1{{\bf #1}}
\def\roman#1{{\rm #1}}
\parskip=\smallskipamount		
\def\quote#1{{\vskip 12pt \moveright 40pt \vbox{\hsize=5truein
    \baselineskip=14pt \noindent #1}}}
\def\boxit#1{\vbox{\hrule\hbox{\vrule\kern3pt         
     \vbox{\kern3pt#1\kern3pt}\kern3pt\vrule}\hrule}} 

\def\fr#1/#2{{\textstyle{#1\over#2}}} 
\def\frac#1/#2{\leavevmode\kern.1em\raise.5ex\hbox{$\scriptstyle #1$}
\kern-.1em/\kern-.15em\lower.25ex\hbox{$\scriptstyle #2$}}
%
%

%

%
%
%
%
%


\newcount\ftnumber
\newdimen\ftparindent\ftparindent\parindent
\long\def\ft#1{\ignorespaces\global\advance\ftnumber by 1
          {\baselineskip\normalbaselineskip
	   \parindent\ftparindent	
           \footnote{$^{\the\ftnumber}$}{\ignorespaces#1}}}
%
\long\def\nft#1{\ignorespaces\global\advance\ftnumber by 1
	\baselineskip\normalbaselineskip
	\footnote{{\sevenrm\raise0.8em\hbox{\the\ftnumber}}}{\ignorespaces#1}}

\long\def\kft#1{\ignorespaces
	{\baselineskip\normalbaselineskip
	 \parindent\ftparindent
	 \vfootnote{$^{\the\ftnumber}$}{#1}}}
%
%
\def\ftname(#1){\expandafter\xdef\csname #1\endcsname{\the\ftnumber}}
%
%
%
%
%
%
%
%
%
\newlinechar=`\^^J%
\newcount\ranfoo%
\ranfoo=\day\multiply\ranfoo\deadcycles%
\advance\ranfoo\pagetotal\advance\ranfoo\time%
\edef\tempenfilnam{/tmp/tex\the\ranfoo.tex}%
\newwrite\tempenfile	
\newwrite\tempenfile%
\newread\tempenrfile
\newread\tempenrfile%
\newcount\ennumber\ennumber=0%
\long\def\endnote#1{%
	\newlinechar=`\^^J%
	\ifnum\number\ennumber=0\immediate\openout\tempenfile=\tempenfilnam\fi%
	\global\advance\ennumber1%
	\immediate\write\tempenfile{#1}%
}%
\newif\ifnotend%
\def\endnotes{%
	\catcode`@=11
	\immediate\closeout\tempenfile%
	\immediate\openin\tempenrfile=\tempenfilnam%
	\loop%
		\ifeof\tempenrfile\notendfalse\else\notendtrue%
			\immediate\read\tempenrfile to \foobie%
			\foobie%
		\fi%
	\ifnotend\repeat%
	\catcode`@=11
}%
\long\def\prlnote#1{%
	\endnote{\par$^{\the\ennumber}$ #1}%
	$^{\the\ennumber}$
}%
%
%
%
%
%
%

\font\tiny = cmr5 scaled 500
\newcount\eqnumber
\def\eq(#1.#2){%
    \ifx\DRAFT\undefined\def\DRAFT{0}\fi	
    \global\advance\eqnumber by 1%
    \expandafter\xdef\csname !#1#2\endcsname{\the\eqnumber}%
    \ifnum\number\DRAFT>0%
	\setbox0=\hbox{\tiny #2}%
	\wd0=0pt%
	\eqno({\offinterlineskip
	  \vtop{\hbox{#1.\the\eqnumber}\vskip1.5pt\box0}})%
    \else%
	\eqno({\rm #1}.\the\eqnumber)%
    \fi%
}
\def\silenteq(#1.#2){%
    \global\advance\eqnumber by 1%
    \global\def\EqTempSecName{#1}%
    \expandafter\xdef\csname !#1#2\endcsname{\the\eqnumber}%
}
\def\alteq(#1){%
    \eqno({\rm \EqTempSecName.\the\eqnumber#1})%
}
\def\(#1.#2){(#1.\csname !#1#2\endcsname)}
\def\altq(#1.#2@#3){({\rm#1.\csname !#1#2\endcsname#3})}

\def\nexteqn(#1+#2){{%
\count0=\the\eqnumber
\advance\count0 #2
(#1.\the\count0)
}}
%
\newbox\FatInternalVariable		
\def\fat#1{{
\kern-.30em
\hbox{					
\setbox\FatInternalVariable=\hbox{#1}
\unskip
\unhcopy\FatInternalVariable
\kern-\wd\FatInternalVariable
\kern 0.015 em
\unhcopy\FatInternalVariable
\kern-\wd\FatInternalVariable
\kern 0.030 em
\unhbox\FatInternalVariable
}
\kern-.33em
}}
%
%
%
%
%
%
%
\font\eightrm=cmr8 \font\eighti=cmmi8 \font\eightsy=cmsy8
\font\eightex=cmex10 scaled 800 
\font\eightit=cmti8 \font\eightsl=cmsl8 \font\eighttt=cmtt8
\font\eightbf=cmbx8
\font\sixrm=cmr7 \font\sixi=cmmi6 \font\sixsy=cmsy6 \font\sixbf=cmbx6
\font\sixex=cmex10 scaled 600 
\newskip\ttglue
\def\eightpoint{\def\rm{\fam0\eightrm}\def\currentpointsize{8}%
  \textfont0=\eightrm\scriptfont0=\sixrm\scriptscriptfont0=\fiverm%
  \textfont1=\eighti\scriptfont1=\sixi\scriptscriptfont1=\fivei%
  \textfont2=\eightsy\scriptfont2=\sixsy\scriptscriptfont2=\fivesy%
  \textfont3=\eightex\scriptfont3=\sixex\scriptscriptfont3=\sixex%
  \textfont\itfam=\eightit \def\it{\fam\itfam\eightit}%
  \textfont\slfam=\eightsl \def\sl{\fam\slfam\eightsl}%
  \textfont\ttfam=\eighttt \def\tt{\fam\ttfam\eighttt}%
  \textfont\bffam=\eightbf \def\bf{\fam\bffam\eightbf}%
  \scriptfont\bffam=\sixbf%
  \scriptscriptfont\bffam=\fivebf%
  \tt\ttglue=.5em plus.25em minus.15em%
  \normalbaselineskip=9pt%
  \setbox\strutbox=\hbox{\vrule height7pt depth2pt width0pt}%
  \ifx\bfitten\undefined\relax\else\let\bfit\bfitten\fi%
  \let\sc=\sixrm\let\big=\eightbig\normalbaselines\rm}%
\def\twelvepoint{\def\currentpointsize{12}%
\font\twelverm=cmr12 \font\twelvei=cmmi12 \font\twelvesy=cmsy10 scaled 1200
\font\twelveex=cmex10 scaled 1200 
\font\twelveit=cmti12 \font\twelvesl=cmsl12 \font\twelvett=cmtt12
\font\twelvebf=cmbx12
\def\twelvescriptscale{960}%
\def\twelvescriptscriptscale{800}%
\font\twrmsc=cmr10 scaled\twelvescriptscale%
\font\twrmscsc=cmr8 scaled\twelvescriptscriptscale%
\font\twisc=cmmi10 scaled\twelvescriptscale%
\font\twiscsc=cmmi8 scaled\twelvescriptscriptscale%
\font\twsysc=cmsy10 scaled\twelvescriptscale%
\font\twsyscsc=cmsy8 scaled\twelvescriptscriptscale%
\font\twexsc=cmex10 scaled\twelvescriptscale%
\font\twexscsc=cmex8 scaled\twelvescriptscriptscale%
  \def\rm{\fam0\twelverm}%
  \textfont0=\twelverm\scriptfont0=\twrmsc\scriptscriptfont0=\twrmscsc%
  \textfont1=\twelvei\scriptfont1=\twisc\scriptscriptfont1=\twiscsc%
  \textfont2=\twelvesy\scriptfont2=\twsysc\scriptscriptfont2=\twsyscsc%
  \textfont3=\twelveex\scriptfont3=\twexsc\scriptscriptfont3=\twexscsc%
  \textfont\itfam=\twelveit \def\it{\fam\itfam\twelveit}%
  \textfont\slfam=\twelvesl \def\sl{\fam\slfam\twelvesl}%
  \textfont\ttfam=\twelvett \def\tt{\fam\ttfam\twelvett}%
  \textfont\bffam=\twelvebf \def\bf{\fam\bffam\twelvebf}%
  \scriptfont\bffam=\tenbf%
  \scriptscriptfont\bffam=\eightbf%
  \tt\ttglue=.75em plus.37em minus.22em%
  \normalbaselineskip=14pt%
  \setbox\strutbox=\hbox{\vrule height10.5pt depth3pt width0pt}%
  \ifx\bfittwelve\undefined\relax\else\let\bfit\bfittwelve\fi%
  \let\sc=\sixrm\let\big=\twelvebig\normalbaselines\rm}
\def\tenpoint{\def\rm{\fam0\tenrm}\def\currentpointsize{10}%
  \textfont0=\tenrm\scriptfont0=\eightrm\scriptscriptfont0=\sixrm%
  \textfont1=\teni\scriptfont1=\eighti\scriptscriptfont1=\sixi%
  \textfont2=\tensy\scriptfont2=\eightsy\scriptscriptfont2=\sixsy%
  \textfont3=\tenex\scriptfont3=\eightex\scriptscriptfont3=\sixex%
  \textfont\itfam=\tenit \def\it{\fam\itfam\tenit}%
  \textfont\slfam=\tensl \def\sl{\fam\slfam\tensl}%
  \textfont\ttfam=\tentt \def\tt{\fam\ttfam\tentt}%
  \textfont\bffam=\tenbf \def\bf{\fam\bffam\tenbf}%
  \scriptfont\bffam=\eightbf%
  \scriptscriptfont\bffam=\sixbf%
  \tt\ttglue=.625em plus.31em minus.185em%
  \normalbaselineskip=11.5pt%
  \setbox\strutbox=\hbox{\vrule height8.75pt depth2.5pt width0pt}%
  \ifx\bfitten\undefined\relax\else\let\bfit\bfitten\fi%
  \let\sc=\sixrm\let\big=\tenbig\normalbaselines\rm}
\font\elevenrm=cmr10 scaled 1100\font\eleveni=cmmi10 scaled 1100%
\font\elevensy=cmsy10 scaled 1100
\font\elevenex=cmex10 scaled 1100 
\font\elevenit=cmti10 scaled 1100
\font\elevensl=cmsl10 scaled 1100
\font\eleventt=cmtt10 scaled 1100
\font\elevenbf=cmbx10 scaled 1100
\font\ninerm=cmr9 
\font\ninebf=cmbx9
\font\ninei=cmmi9\font\ninesy=cmsy9\font\nineex=cmex9
\def\elevenpoint{\def\rm{\fam0\elevenrm}\def\currentpointsize{11}%
  \textfont0=\elevenrm\scriptfont0=\ninerm\scriptscriptfont0=\eightrm%
  \textfont1=\eleveni\scriptfont1=\ninei\scriptscriptfont1=\eighti%
  \textfont2=\elevensy\scriptfont2=\ninesy\scriptscriptfont2=\eightsy%
  \textfont3=\elevenex\scriptfont3=\nineex\scriptscriptfont3=\eightex%
  \textfont\itfam=\elevenit \def\it{\fam\itfam\elevenit}%
  \textfont\slfam=\elevensl \def\sl{\fam\slfam\elevensl}%
  \textfont\ttfam=\eleventt \def\tt{\fam\ttfam\eleventt}%
  \textfont\bffam=\elevenbf \def\bf{\fam\bffam\elevenbf}%
  \scriptfont\bffam=\ninebf%
  \scriptscriptfont\bffam=\eightbf%
  \tt\ttglue=.6875em plus.34em minus.2025em%
  \normalbaselineskip=12.75pt%
  \setbox\strutbox=\hbox{\vrule height9.625pt depth2.75pt width0pt}%
  \ifx\bfiteleven\undefined\relax\else\let\bfit\bfiteleven\fi%
  \let\sc=\sixrm\let\big=\elevenbig\normalbaselines\rm}
%
%
\font\fourteenrm=cmr12 scaled 1167\font\fourteeni=cmmi12 scaled 1167\font\fourteensy=cmsy10 scaled 1400
\font\fourteenex=cmex10 scaled 1400 
\font\fourteenit=cmti12 scaled 1167 \font\fourteensl=cmsl12  scaled 1167\font\fourteentt=cmtt12 scaled 1167
\font\fourteenbf=cmbx12 scaled 1167
\font\twrmsc=cmr10 scaled1120%
\font\twrmscsc=cmr8 scaled1050%
\font\twisc=cmmi10 scaled1120%
\font\twiscsc=cmmi8 scaled1050%
\font\twsysc=cmsy10 scaled1120%
\font\twsyscsc=cmsy8 scaled1050%
\font\twexsc=cmex10 scaled1120%
\font\twexscsc=cmex8 scaled1050%
\font\twelvebf=cmbx12%
\def\fourteenpoint{\def\rm{\fam0\fourteenrm}\def\currentpointsize{14}%
  \textfont0=\fourteenrm\scriptfont0=\twrmsc\scriptscriptfont0=\twrmscsc%
  \textfont1=\fourteeni\scriptfont1=\twisc\scriptscriptfont1=\twiscsc%
  \textfont2=\fourteensy\scriptfont2=\twsysc\scriptscriptfont2=\twsyscsc%
  \textfont3=\fourteenex\scriptfont3=\twexsc\scriptscriptfont3=\twexscsc%
  \textfont\itfam=\fourteenit \def\it{\fam\itfam\fourteenit}%
  \textfont\slfam=\fourteensl \def\sl{\fam\slfam\fourteensl}%
  \textfont\ttfam=\fourteentt \def\tt{\fam\ttfam\fourteentt}%
  \textfont\bffam=\fourteenbf \def\bf{\fam\bffam\fourteenbf}%
  \scriptfont\bffam=\twelvebf%
  \scriptscriptfont\bffam=\tenbf%
  \tt\ttglue=.875em plus.43em minus.26em%
  \normalbaselineskip=16pt%
  \setbox\strutbox=\hbox{\vrule height12.3pt depth3.5pt width0pt}%
  \ifx\bfitfourteen\undefined\relax\else\let\bfit\bfitfourteen\fi%
  \let\sc=\twelverm\let\big=\fourteenbig\normalbaselines\rm}
%
\font\nineit=cmti9\font\ninesl=cmsl9\font\ninett=cmtt9
\font\sevenrm=cmr7\font\seveni=cmmi7\font\sevensy=cmsy7
\font\sevenex=cmex10 scaled 700
\font\sevenbf=cmbx7
\def\ninepoint{\def\rm{\fam0\ninerm}\def\currentpointsize{9}%
  \textfont0=\ninerm\scriptfont0=\sevenrm\scriptscriptfont0=\fiverm%
  \textfont1=\ninei\scriptfont1=\seveni\scriptscriptfont1=\fivei%
  \textfont2=\ninesy\scriptfont2=\sevensy\scriptscriptfont2=\fivesy%
  \textfont3=\nineex\scriptfont3=\sevenex\scriptscriptfont3=\sevenex%
  \textfont\itfam=\nineit \def\it{\fam\itfam\nineit}%
  \textfont\slfam=\ninesl \def\sl{\fam\slfam\ninesl}%
  \textfont\ttfam=\ninett \def\tt{\fam\ttfam\ninett}%
  \textfont\bffam=\ninebf \def\bf{\fam\bffam\ninebf}%
  \scriptfont\bffam=\sevenbf%
  \scriptscriptfont\bffam=\fivebf%
  \tt\ttglue=.5625em plus.279em minus.1665em%
  \normalbaselineskip=10.35pt%
  \setbox\strutbox=\hbox{\vrule height7.875pt depth1.8pt width0pt}%
  \ifx\bfitten\undefined\relax\else\let\bfit\bfitten\fi%
  \let\sc=\sevenrm\let\big=\ninebig\normalbaselines\rm}%
%
%
%
\font\sixteenrm=cmr12 scaled 1333\font\sixteeni=cmmi12 scaled 1333\font\sixteensy=cmsy10 scaled 1600
\font\sixteenex=cmex10 scaled 1600 
\font\sixteenit=cmti12 scaled 1333 \font\sixteensl=cmsl12  scaled 1333\font\sixteentt=cmtt12 scaled 1333
\font\sixteenbf=cmbx12 scaled 1333
\font\twrmsc=cmr10 scaled1280%
\font\twrmscsc=cmr8 scaled1200%
\font\twisc=cmmi10 scaled1280%
\font\twiscsc=cmmi8 scaled1200%
\font\twsysc=cmsy10 scaled1280%
\font\twsyscsc=cmsy8 scaled1200%
\font\twexsc=cmex10 scaled1280%
\font\twexscsc=cmex8 scaled1200%
\font\fourteenbf=cmbx12 scaled1167%
\def\sixteenpoint{\def\rm{\fam0\sixteenrm}\def\currentpointsize{16}%
  \textfont0=\sixteenrm\scriptfont0=\twrmsc\scriptscriptfont0=\twrmscsc%
  \textfont1=\sixteeni\scriptfont1=\twisc\scriptscriptfont1=\twiscsc%
  \textfont2=\sixteensy\scriptfont2=\twsysc\scriptscriptfont2=\twsyscsc%
  \textfont3=\sixteenex\scriptfont3=\twexsc\scriptscriptfont3=\twexscsc%
  \textfont\itfam=\sixteenit \def\it{\fam\itfam\sixteenit}%
  \textfont\slfam=\sixteensl \def\sl{\fam\slfam\sixteensl}%
  \textfont\ttfam=\sixteentt \def\tt{\fam\ttfam\sixteentt}%
  \textfont\bffam=\sixteenbf \def\bf{\fam\bffam\sixteenbf}%
  \scriptfont\bffam=\fourteenbf%
  \scriptscriptfont\bffam=\tenbf%
  \tt\ttglue=1em plus.491em minus.297em%
  \normalbaselineskip=18pt%
  \setbox\strutbox=\hbox{\vrule height14.1pt depth4pt width0pt}%
  \ifx\bfitsixteen\undefined\relax\else\let\bfit\bfitsixteen\fi%
  \let\sc=\twelverm\let\big=\sixteenbig\normalbaselines\rm}
%

%
\def\bold#1{{\bf#1}}


\def\header#1{\bigskip\centerline{\bf#1}\nobreak}
\def\subsection#1#2{\header{#1. #2}}

\long\def\defbox#1=#2{\newbox#1\setbox#1=#2} 
\def\defdimen#1=#2{\newdimen#1#1=#2}



%
%
%
\long\def\VV#1#2{{\hbox{%
\newbox\VVA\newbox\VVB%
\setbox\VVA=\hbox{#2}%
\newdimen\VVC\VVC=\wd\VVA%
\setbox\VVB=\hbox{\vbox to#1{\vfil\hsize=\VVC\hbox{\unhbox\VVA}\vfil}}%
\wd\VVB=\VVC%
\divide\VVC2
\unhbox\VVB}}}
%

\def\={\equiv}

\def\seq(#1){\eq(\SEC.#1)}
\def\seqr(#1){\(\SEC.#1)}

\defdimen\fakeIPleft=0pt
\def\fakeIPinsert{\relax}%
\long\def\fakeIP#1#2{{\defdimen\oldleft=\leftskip%
\defdimen\oldparindent=\the\parindent%
\long\def\fakeft##1{{\leftskip\oldleft\parindent=\the\oldparindent\ft{##1}}}%
\defbox\bletch=\hbox{\noindent#1\hbox to \IPspace{}\hfil}%
\ifdim\fakeIPleft<\wd\bletch\relax\fakeIPleft=\wd\bletch\relax\fi%
\global\leftskip=\fakeIPleft\noindent%
\llap{\hbox to\leftskip{#1\hfil}}%
\fakeIPinsert%
\noindent\parindent=0pt%
\parindent=0pt%
\ignorespaces%
\def\fakeIPkludge{%
\displayindent=\leftskip\advance\displaywidth-\displayindent}%
{\noindent\ignorespaces#2}\par\global\leftskip=\oldleft}} 
%
\long\def\cstok#1{\def\hsp{\hphantom{.}}
	\def\vsp{\kern2pt}
	\leavevmode\hsp\hbox{\vrule\vtop{\vbox{\hrule\kern1pt\vsp%
	\hbox{\vphantom{/}\hsp{#1}\hsp}}%
	\kern1pt\vsp\hrule}\vrule}\hsp}
%
%

%
%
%
\font\cmbxti = cmbxti10	
\font\cmmib  = cmmib10	
\font\cmbxtitwelve = cmbxti10 scaled 1200
\font\cmmibtwelve = cmmib10 scaled 1200
\font\cmbxtifourteen = cmbxti10 scaled 1400
\font\cmmibfourteen = cmmib10 scaled 1400
\newfam\mibtenfam \textfont\mibtenfam=\cmmib \scriptfont\mibtenfam=\cmmib
	\scriptscriptfont\mibtenfam=\cmmib
\newfam\mibtwelvefam\textfont\mibtwelvefam=\cmmibtwelve
	\scriptfont\mibtwelvefam=\cmmibtwelve
	\scriptscriptfont\mibtwelvefam=\cmmibtwelve
\newfam\mibfourteenfam\textfont\mibfourteenfam=\cmmibfourteen
	\scriptfont\mibfourteenfam=\cmmibfourteen
	\scriptscriptfont\mibfourteenfam=\cmmibfourteen
%
\def\bfitten{\fam\mibtenfam\cmbxti}
\def\bfittwelve{\fam\mibtwelvefam\cmbxtitwelve}
\def\bfitfourteen{\fam\mibfourteenfam\cmbxtifourteen}
\let\bfit\bfitten
%
\mathchardef\alpha="710B\mathchardef\beta="710C\mathchardef\gamma="710D
\mathchardef\delta="710E\mathchardef\epsilon="710F\mathchardef\zeta="7110
\mathchardef\eta="7111\mathchardef\theta="7112\mathchardef\iota="7113
\mathchardef\kappa="7114\mathchardef\lambda="7115\mathchardef\mu="7116
\mathchardef\nu="7117\mathchardef\xi="7118\mathchardef\pi="7119
\mathchardef\rho="711A\mathchardef\sigma="711B\mathchardef\tau="711C
\mathchardef\upsilon="711D\mathchardef\phi="711E\mathchardef\chi="711F
\mathchardef\psi="7120\mathchardef\omega="7121\mathchardef\varepsilon="7122
\mathchardef\vartheta="7123\mathchardef\varpi="7124\mathchardef\varrho="7125
\mathchardef\varsigma="7126\mathchardef\varphi="7127
%
%
%
%
\newcount\captionnumber\captionnumber0
\long\def\caption#1{\ninepoint\baselineskip16pt#1\par%
\global\advance\captionnumber1\relax%
\expandafter\gdef\csname!caption\the\captionnumber\endcsname{#1}}
\def\recaption#1{\expandafter\csname!caption#1\endcsname\par}
\def\header#1{\medskip\centerline{\bf#1}\nobreak\smallskip\nobreak}

\def\ens#1{\expandafter\xdef\csname!endnote#1\endcsname{\the\ennumber}}
\def\enr#1{\csname!endnote#1\endcsname}
\long\def\prlnotez#1{\endnote{\par$^{\the\ennumber}$ #1}}

\font\msxm msam10\relax
\textfont"E\msxm
\mathchardef\geo"3E26
\mathchardef\leo"3E2E
\twelvepoint
\baselineskip22pt

%
%
%
%

\font\tiny = cmr5 scaled 500
\newcount\eqnumber
\def\eq(#1){%
    \ifx\DRAFT\undefined\def\DRAFT{0}\fi	
    \global\advance\eqnumber by 1%
    \expandafter\xdef\csname !#1\endcsname{\the\eqnumber}%
    \ifnum\number\DRAFT>0%
	\setbox0=\hbox{\tiny #1}%
	\wd0=0pt%
	\eqno({\offinterlineskip
	  \vtop{\hbox{\the\eqnumber}\vskip1.5pt\box0}})%
    \else%
	\eqno(\the\eqnumber)%
    \fi%
}
\def\silenteq(#1){%
    \global\advance\eqnumber by 1%
    \expandafter\xdef\csname !#1\endcsname{\the\eqnumber}%
}
\def\alteq(#1){%
    \eqno({\rm\the\eqnumber#1})%
}
\def\(#1){(\csname !#1\endcsname)}
\def\altq(#1@#2){({\rm\csname !#1\endcsname#2})}

\def\nexteqn(#1){{%
\count0=\the\eqnumber
\advance\count0 #1
(\the\count0)
}}
%
%
%
%
%
%
%
%
%
%
\font\eightrm=cmr8 \font\eighti=cmmi8 \font\eightsy=cmsy8
\font\eightex=cmex10 scaled 800 
\font\eightit=cmti8 \font\eightsl=cmsl8 \font\eighttt=cmtt8
\font\eightbf=cmbx8
\font\sixrm=cmr6 \font\sixi=cmmi6 \font\sixsy=cmsy6 \font\sixbf=cmbx6
\font\sixex=cmex10 scaled 600 
\newskip\ttglue
\def\eightpoint{\def\rm{\fam0\eightrm}%
  \textfont0=\eightrm\scriptfont0=\sixrm\scriptscriptfont0=\fiverm%
  \textfont1=\eighti\scriptfont1=\sixi\scriptscriptfont1=\fivei%
  \textfont2=\eightsy\scriptfont2=\sixsy\scriptscriptfont2=\fivesy%
  \textfont3=\eightex\scriptfont3=\sixex\scriptscriptfont3=\sixex%
  \textfont\itfam=\eightit \def\it{\fam\itfam\eightit}%
  \textfont\slfam=\eightsl \def\sl{\fam\slfam\eightsl}%
  \textfont\ttfam=\eighttt \def\tt{\fam\ttfam\eighttt}%
  \textfont\bffam=\eightbf \def\bf{\fam\bffam\eightbf}%
  \scriptfont\bffam=\sixbf%
  \scriptscriptfont\bffam=\fivebf%
  \tt\ttglue=.5em plus.25em minus.15em%
  \normalbaselineskip=9pt%
  \setbox\strutbox=\hbox{\vrule height7pt depth2pt width0pt}%
  \ifx\bfitten\undefined\relax\else\let\bfit\bfitten\fi%
  \let\sc=\sixrm\let\big=\eightbig\normalbaselines\rm}%
\def\twelvepoint{%
\font\twelverm=cmr12 \font\twelvei=cmmi12 \font\twelvesy=cmsy10 scaled 1200
\font\twelveex=cmex10 scaled 1200 
\font\twelveit=cmti12 \font\twelvesl=cmsl12 \font\twelvett=cmtt12
\font\twelvebf=cmbx12
\def\twelvescriptscale{960}%
\def\twelvescriptscriptscale{800}%
\font\twrmsc=cmr10 scaled\twelvescriptscale%
\font\twrmscsc=cmr8 scaled\twelvescriptscriptscale%
\font\twisc=cmmi10 scaled\twelvescriptscale%
\font\twiscsc=cmmi8 scaled\twelvescriptscriptscale%
\font\twsysc=cmsy10 scaled\twelvescriptscale%
\font\twsyscsc=cmsy8 scaled\twelvescriptscriptscale%
\font\twexsc=cmex10 scaled\twelvescriptscale%
\font\twexscsc=cmex8 scaled\twelvescriptscriptscale%
  \def\rm{\fam0\twelverm}%
  \textfont0=\twelverm\scriptfont0=\twrmsc\scriptscriptfont0=\twrmscsc%
  \textfont1=\twelvei\scriptfont1=\twisc\scriptscriptfont1=\twiscsc%
  \textfont2=\twelvesy\scriptfont2=\twsysc\scriptscriptfont2=\twsyscsc%
  \textfont3=\twelveex\scriptfont3=\twexsc\scriptscriptfont3=\twexscsc%
  \textfont\itfam=\twelveit \def\it{\fam\itfam\twelveit}%
  \textfont\slfam=\twelvesl \def\sl{\fam\slfam\twelvesl}%
  \textfont\ttfam=\twelvett \def\tt{\fam\ttfam\twelvett}%
  \textfont\bffam=\twelvebf \def\bf{\fam\bffam\twelvebf}%
  \scriptfont\bffam=\tenbf%
  \scriptscriptfont\bffam=\eightbf%
  \tt\ttglue=.75em plus.37em minus.22em%
  \normalbaselineskip=14pt%
  \setbox\strutbox=\hbox{\vrule height10.5pt depth3pt width0pt}%
  \ifx\bfittwelve\undefined\relax\else\let\bfit\bfittwelve\fi%
  \let\sc=\sixrm\let\big=\twelvebig\normalbaselines\rm}
\def\tenpoint{\def\rm{\fam0\tenrm}%
  \textfont0=\tenrm\scriptfont0=\eightrm\scriptscriptfont0=\sixrm%
  \textfont1=\teni\scriptfont1=\eighti\scriptscriptfont1=\sixi%
  \textfont2=\tensy\scriptfont2=\eightsy\scriptscriptfont2=\sixsy%
  \textfont3=\tenex\scriptfont3=\eightex\scriptscriptfont3=\sixex%
  \textfont\itfam=\tenit \def\it{\fam\itfam\tenit}%
  \textfont\slfam=\tensl \def\sl{\fam\slfam\tensl}%
  \textfont\ttfam=\tentt \def\tt{\fam\ttfam\tentt}%
  \textfont\bffam=\tenbf \def\bf{\fam\bffam\tenbf}%
  \scriptfont\bffam=\eightbf%
  \scriptscriptfont\bffam=\sixbf%
  \tt\ttglue=.625em plus.31em minus.185em%
  \normalbaselineskip=11.5pt%
  \setbox\strutbox=\hbox{\vrule height8.75pt depth2.5pt width0pt}%
  \ifx\bfitten\undefined\relax\else\let\bfit\bfitten\fi%
  \let\sc=\sixrm\let\big=\tenbig\normalbaselines\rm}
\font\elevenrm=cmr10 scaled 1100\font\eleveni=cmmi10 scaled 1100%
\font\elevensy=cmsy10 scaled 1100
\font\elevenex=cmex10 scaled 1100 
\font\elevenit=cmti10 scaled 1100
\font\elevensl=cmsl10 scaled 1100
\font\eleventt=cmtt10 scaled 1100
\font\elevenbf=cmbx10 scaled 1100
\font\ninerm=cmr9 
\font\ninebf=cmbx9
\font\ninei=cmmi9\font\ninesy=cmsy9\font\nineex=cmex9
\def\elevenpoint{\def\rm{\fam0\elevenrm}%
  \textfont0=\elevenrm\scriptfont0=\ninerm\scriptscriptfont0=\eightrm%
  \textfont1=\eleveni\scriptfont1=\ninei\scriptscriptfont1=\eighti%
  \textfont2=\elevensy\scriptfont2=\ninesy\scriptscriptfont2=\eightsy%
  \textfont3=\elevenex\scriptfont3=\nineex\scriptscriptfont3=\eightex%
  \textfont\itfam=\elevenit \def\it{\fam\itfam\elevenit}%
  \textfont\slfam=\elevensl \def\sl{\fam\slfam\elevensl}%
  \textfont\ttfam=\eleventt \def\tt{\fam\ttfam\eleventt}%
  \textfont\bffam=\elevenbf \def\bf{\fam\bffam\elevenbf}%
  \scriptfont\bffam=\ninebf%
  \scriptscriptfont\bffam=\eightbf%
  \tt\ttglue=.6875em plus.34em minus.2025em%
  \normalbaselineskip=12.75pt%
  \setbox\strutbox=\hbox{\vrule height9.625pt depth2.75pt width0pt}%
  \ifx\bfiteleven\undefined\relax\else\let\bfit\bfiteleven\fi%
  \let\sc=\sixrm\let\big=\elevenbig\normalbaselines\rm}
%
%
\font\fourteenrm=cmr12 scaled 1167\font\fourteeni=cmmi12 scaled 1167\font\fourteensy=cmsy10 scaled 1400
\font\fourteenex=cmex10 scaled 1400 
\font\fourteenit=cmti12 scaled 1167 \font\fourteensl=cmsl12  scaled 1167\font\fourteentt=cmtt12 scaled 1167
\font\fourteenbf=cmbx12 scaled 1167
\font\twrmsc=cmr10 scaled1120%
\font\twrmscsc=cmr8 scaled1050%
\font\twisc=cmmi10 scaled1120%
\font\twiscsc=cmmi8 scaled1050%
\font\twsysc=cmsy10 scaled1120%
\font\twsyscsc=cmsy8 scaled1050%
\font\twexsc=cmex10 scaled1120%
\font\twexscsc=cmex8 scaled1050%
\font\twelvebf=cmbx12%
\def\fourteenpoint{\def\rm{\fam0\fourteenrm}%
  \textfont0=\fourteenrm\scriptfont0=\twrmsc\scriptscriptfont0=\twrmscsc%
  \textfont1=\fourteeni\scriptfont1=\twisc\scriptscriptfont1=\twiscsc%
  \textfont2=\fourteensy\scriptfont2=\twsysc\scriptscriptfont2=\twsyscsc%
  \textfont3=\fourteenex\scriptfont3=\twexsc\scriptscriptfont3=\twexscsc%
  \textfont\itfam=\fourteenit \def\it{\fam\itfam\fourteenit}%
  \textfont\slfam=\fourteensl \def\sl{\fam\slfam\fourteensl}%
  \textfont\ttfam=\fourteentt \def\tt{\fam\ttfam\fourteentt}%
  \textfont\bffam=\fourteenbf \def\bf{\fam\bffam\fourteenbf}%
  \scriptfont\bffam=\twelvebf%
  \scriptscriptfont\bffam=\tenbf%
  \tt\ttglue=.875em plus.43em minus.26em%
  \normalbaselineskip=16pt%
  \setbox\strutbox=\hbox{\vrule height12.3pt depth3.5pt width0pt}%
  \ifx\bfitfourteen\undefined\relax\else\let\bfit\bfitfourteen\fi%
  \let\sc=\twelverm\let\big=\fourteenbig\normalbaselines\rm}
%
\font\nineit=cmti9\font\ninesl=cmsl9\font\ninett=cmtt9
\font\sevenrm=cmr7\font\seveni=cmmi7\font\sevensy=cmsy7
\font\sevenex=cmex10 scaled 700
\font\sevenbf=cmbx7
\def\ninepoint{\def\rm{\fam0\ninerm}%
  \textfont0=\ninerm\scriptfont0=\sevenrm\scriptscriptfont0=\fiverm%
  \textfont1=\ninei\scriptfont1=\seveni\scriptscriptfont1=\fivei%
  \textfont2=\ninesy\scriptfont2=\sevensy\scriptscriptfont2=\fivesy%
  \textfont3=\nineex\scriptfont3=\sevenex\scriptscriptfont3=\sevenex%
  \textfont\itfam=\nineit \def\it{\fam\itfam\nineit}%
  \textfont\slfam=\ninesl \def\sl{\fam\slfam\ninesl}%
  \textfont\ttfam=\ninett \def\tt{\fam\ttfam\ninett}%
  \textfont\bffam=\ninebf \def\bf{\fam\bffam\ninebf}%
  \scriptfont\bffam=\sevenbf%
  \scriptscriptfont\bffam=\fivebf%
  \tt\ttglue=.5625em plus.279em minus.1665em%
  \normalbaselineskip=10.35pt%
  \setbox\strutbox=\hbox{\vrule height7.875pt depth1.8pt width0pt}%
  \ifx\bfitten\undefined\relax\else\let\bfit\bfitten\fi%
  \let\sc=\sevenrm\let\big=\ninebig\normalbaselines\rm}%
%
%
%
\newbox\texboxbox%
\long\def\texbox#1{{%
\setbox\texboxbox\hbox{#1}%
\hbox{\vrule width0.4pt\vbox{%
\hrule height0.4pt\hbox{%
\vbox{%
\hbox{%
\unhcopy\texboxbox}%
\kern-\the\dp\texboxbox\hrule height0.15pt%
\kern\the\dp\texboxbox%
}}%
\hrule height0.4pt}%
\vrule width0.4pt}%
}}%
%
\long\def\boxx#1{{%
\setbox0\hbox{#1}%
\dimen0\dp0%
\advance\dimen0 1.5pt%
\hbox{\vrule{\lower\dimen0\vbox{\hrule\hbox spread 3pt{\hfil\vbox spread 3pt{\vfil\box0\vfil}\hfil}\hrule}\vrule}}%
}}

\input eplain
\input epsf
\def\DRAFT{0}
\edef\cite{\the\catcode`@}%
\catcode`@ = 11
\let\@oldatcatcode = \cite
\chardef\@letter = 11
\chardef\@other = 12
%
%
%
%
\def\@innerdef#1#2{\edef#1{\expandafter\noexpand\csname #2\endcsname}}%
%
%
\@innerdef\@innernewcount{newcount}%
\@innerdef\@innernewdimen{newdimen}%
\@innerdef\@innernewif{newif}%
\@innerdef\@innernewwrite{newwrite}%
%
%
%
\def\@gobble#1{}%
%
%
%
\ifx\inputlineno\@undefined
   \let\@linenumber = \empty 
\else
   \def\@linenumber{\the\inputlineno:\space}%
\fi
%
%
%
\def\@futurenonspacelet#1{\def\cs{#1}%
   \afterassignment\@stepone\let\@nexttoken=
}%
\begingroup 
\def\\{\global\let\@stoken= }%
\\ 
\endgroup
\def\@stepone{\expandafter\futurelet\cs\@steptwo}%
\def\@steptwo{\expandafter\ifx\cs\@stoken\let\@@next=\@stepthree
   \else\let\@@next=\@nexttoken\fi \@@next}%
\def\@stepthree{\afterassignment\@stepone\let\@@next= }%
%
%
%
\def\@getoptionalarg#1{%
   \let\@optionaltemp = #1%
   \let\@optionalnext = \relax
   \@futurenonspacelet\@optionalnext\@bracketcheck
}%
%
%
\def\@bracketcheck{%
   \ifx [\@optionalnext
      \expandafter\@@getoptionalarg
   \else
      \let\@optionalarg = \empty
      \expandafter\@optionaltemp
   \fi
}%
\def\@@getoptionalarg[#1]{%
   \def\@optionalarg{#1}%
   \@optionaltemp
}%
%
%
%
\def\@nnil{\@nil}%
\def\@fornoop#1\@@#2#3{}%
\def\@for#1:=#2\do#3{%
   \edef\@fortmp{#2}%
   \ifx\@fortmp\empty \else
      \expandafter\@forloop#2,\@nil,\@nil\@@#1{#3}%
   \fi
}%
\def\@forloop#1,#2,#3\@@#4#5{\def#4{#1}\ifx #4\@nnil \else
       #5\def#4{#2}\ifx #4\@nnil \else#5\@iforloop #3\@@#4{#5}\fi\fi
}%
\def\@iforloop#1,#2\@@#3#4{\def#3{#1}\ifx #3\@nnil
       \let\@nextwhile=\@fornoop \else
      #4\relax\let\@nextwhile=\@iforloop\fi\@nextwhile#2\@@#3{#4}%
}%
%
%
%
\@innernewif\if@fileexists
\def\@testfileexistence{\@getoptionalarg\@finishtestfileexistence}%
\def\@finishtestfileexistence#1{%
   \begingroup
      \def\extension{#1}%
      \immediate\openin0 =
         \ifx\@optionalarg\empty\jobname\else\@optionalarg\fi
         \ifx\extension\empty \else .#1\fi
         \space
      \ifeof 0
         \global\@fileexistsfalse
      \else
         \global\@fileexiststrue
      \fi
      \immediate\closein0
   \endgroup
}%
%
%
%
%
\def\bibliographystyle#1{%
   \@readauxfile
   \@writeaux{\string\bibstyle{#1}}%
}%
\let\bibstyle = \@gobble
%
%
\let\bblfilebasename = \jobname
\def\bibliography#1{%
   \@readauxfile
   \@writeaux{\string\bibdata{#1}}%
   \@testfileexistence[\bblfilebasename]{bbl}%
   \if@fileexists
      \nobreak
      \@readbblfile
   \fi
}%
\let\bibdata = \@gobble
%
%
\def\nocite#1{%
   \@readauxfile
   \@writeaux{\string\citation{#1}}%
}%
\@innernewif\if@notfirstcitation
%
%
\def\cite{\@getoptionalarg\@cite}%
%
%
\def\@cite#1{%
   \let\@citenotetext = \@optionalarg
   \printcitestart
   \nocite{#1}%
   \@notfirstcitationfalse
   \@for \@citation :=#1\do
   {%
      \expandafter\@onecitation\@citation\@@
   }%
   \ifx\empty\@citenotetext\else
      \printcitenote{\@citenotetext}%
   \fi
   \printcitefinish
}%
\def\@onecitation#1\@@{%
   \if@notfirstcitation
      \printbetweencitations
   \fi
   \expandafter \ifx \csname\@citelabel{#1}\endcsname \relax
      \if@citewarning
         \message{\@linenumber Undefined citation `#1'.}%
      \fi
      \expandafter\gdef\csname\@citelabel{#1}\endcsname{%
         {\tt
            \escapechar = -1
            \nobreak\hskip0pt
            \expandafter\string\csname#1\endcsname
            \nobreak\hskip0pt
         }%
      }%
   \fi
   \csname\@citelabel{#1}\endcsname
   \@notfirstcitationtrue
}%
%
%
\def\@citelabel#1{b@#1}%
%
%
\def\@citedef#1#2{\expandafter\gdef\csname\@citelabel{#1}\endcsname{#2}}%
%
%
%
\def\@readbblfile{%
   \ifx\@itemnum\@undefined
      \@innernewcount\@itemnum
   \fi
   \begingroup
      \ifx\begin\@undefined
         \def\begin##1##2{%
            \setbox0 = \hbox{\biblabelcontents{##2}}%
            \biblabelwidth = \wd0
         }%
         \let\end = \@gobble 
      \fi
      %
      %
      \@itemnum = 0
      \def\bibitem{\@getoptionalarg\@bibitem}%
      \def\@bibitem{%
         \ifx\@optionalarg\empty
            \expandafter\@numberedbibitem
         \else
            \expandafter\@alphabibitem
         \fi
      }%
      \def\@alphabibitem##1{%
         \expandafter \xdef\csname\@citelabel{##1}\endcsname {\@optionalarg}%
         \ifx\biblabelprecontents\@undefined
            \let\biblabelprecontents = \relax
         \fi
         \ifx\biblabelpostcontents\@undefined
            \let\biblabelpostcontents = \hss
         \fi
         \@finishbibitem{##1}%
      }%
      \def\@numberedbibitem##1{%
         \advance\@itemnum by 1
         \expandafter \xdef\csname\@citelabel{##1}\endcsname{\number\@itemnum}%
         \ifx\biblabelprecontents\@undefined
            \let\biblabelprecontents = \hss
         \fi
         \ifx\biblabelpostcontents\@undefined
            \let\biblabelpostcontents = \relax
         \fi
         \@finishbibitem{##1}%
      }%
      \def\@finishbibitem##1{%
         \biblabelprint{\csname\@citelabel{##1}\endcsname}%
         \@writeaux{\string\@citedef{##1}{\csname\@citelabel{##1}\endcsname}}%
         \ignorespaces
      }%
      %
      %
      \let\em = \bblem
      \let\newblock = \bblnewblock
      \let\sc = \bblsc
      \frenchspacing
      \clubpenalty = 4000 \widowpenalty = 4000
      \tolerance = 10000 \hfuzz = .5pt
      \everypar = {\hangindent = \biblabelwidth
                      \advance\hangindent by \biblabelextraspace}%
      \bblrm
      \parskip = 1.5ex plus .5ex minus .5ex
      \biblabelextraspace = .5em
      \bblhook
      \input \bblfilebasename.bbl
   \endgroup
}%
%
%
\@innernewdimen\biblabelwidth
\@innernewdimen\biblabelextraspace
%
%
%
\def\biblabelprint#1{%
   \noindent
   \hbox to \biblabelwidth{%
      \biblabelprecontents
      \biblabelcontents{#1}%
      \biblabelpostcontents
   }%
   \kern\biblabelextraspace
}%
%
%
%
\def\biblabelcontents#1{{\bblrm [#1]}}%
%
%
\def\bblrm{\rm}%
%
%
\def\bblem{\it}%
%
%
\def\bblsc{\ifx\@scfont\@undefined
              \font\@scfont = cmcsc10
           \fi
           \@scfont
}%
%
%
\def\bblnewblock{\hskip .11em plus .33em minus .07em }%
%
%
\let\bblhook = \empty
%
%
%
\def\printcitestart{[}
\def\printcitefinish{]}
\def\printbetweencitations{, }
\def\printcitenote#1{, #1}
%
%
%
\let\citation = \@gobble
%
%
%
\@innernewcount\@numparams
%
%
\def\newcommand#1{%
   \def\@commandname{#1}%
   \@getoptionalarg\@continuenewcommand
}%
%
%
\def\@continuenewcommand{%
   \@numparams = \ifx\@optionalarg\empty 0\else\@optionalarg \fi \relax
   \@newcommand
}%
%
%
\def\@newcommand#1{%
   \def\@startdef{\expandafter\edef\@commandname}%
   \ifnum\@numparams=0
      \let\@paramdef = \empty
   \else
      \ifnum\@numparams>9
         \errmessage{\the\@numparams\space is too many parameters}%
      \else
         \ifnum\@numparams<0
            \errmessage{\the\@numparams\space is too few parameters}%
         \else
            \edef\@paramdef{%
               \ifcase\@numparams
                  \empty  No arguments.
               \or ####1%
               \or ####1####2%
               \or ####1####2####3%
               \or ####1####2####3####4%
               \or ####1####2####3####4####5%
               \or ####1####2####3####4####5####6%
               \or ####1####2####3####4####5####6####7%
               \or ####1####2####3####4####5####6####7####8%
               \or ####1####2####3####4####5####6####7####8####9%
               \fi
            }%
         \fi
      \fi
   \fi
   \expandafter\@startdef\@paramdef{#1}%
}%
%
%
%
%
\def\@readauxfile{%
   \if@auxfiledone \else 
      \global\@auxfiledonetrue
      \@testfileexistence{aux}%
      \if@fileexists
         \begingroup
            \endlinechar = -1
            \catcode`@ = 11
            \input \jobname.aux
         \endgroup
      \else
         \message{\@undefinedmessage}%
         \global\@citewarningfalse
      \fi
      \immediate\openout\@auxfile = \jobname.aux
   \fi
}%
%
%
\newif\if@auxfiledone
\ifx\noauxfile\@undefined \else \@auxfiledonetrue\fi
%
%
%
%
\@innernewwrite\@auxfile
\def\@writeaux#1{\ifx\noauxfile\@undefined\immediate\write\@auxfile{#1}\fi}%
%
%
%
\ifx\@undefinedmessage\@undefined
   \def\@undefinedmessage{No .aux file; I won't give you warnings about
                          undefined citations.}%
\fi
%
%
\@innernewif\if@citewarning
\ifx\noauxfile\@undefined \@citewarningtrue\fi
%
%
%
\catcode`@ = \@oldatcatcode
%
%
%
\catcode`@11

%
\def\ncite@atoi#1{0#1}

\def\ncite#1{%
\def\ncite@cur{}
\innernewcount\ncite@s\ncite@s0
\innernewcount\ncite@c
\innernewcount\ncite@p
\unskip$^\bgroup\,
\for\name:=#1\do{
\nocite{\name}
\xdef\ncite@cur{\expandafter\csname\@citelabel{\name}\endcsname}
\ncite@p\ncite@c%
\ncite@c=\ncite@atoi{\ncite@cur}\relax%
\ifcase\ncite@s
\the\ncite@c
\ncite@s=1%
\or%
\ncite@one%
\or%
\ncite@two%
\fi
}
\ifcase\ncite@s\relax\or\ncite@three\or\ncite@four\fi
\egroup\,$
}

\def\ncite@one{%
\advance\ncite@p1\relax
\ifnum\ncite@p=\ncite@c\relax\ncite@s2\relax\else%
\advance\ncite@p-1\relax%
,\the\ncite@c\relax
\fi%
}

\def\ncite@two{%
\advance\ncite@p1\relax
\ifnum\ncite@p=\ncite@c\relax\else%
\advance\ncite@p-1\relax%
\hbox{--}\the\ncite@c\relax
\ncite@s1\relax%
\fi%
}

\def\ncite@three{%
}

\def\ncite@four{%
\ifnum\ncite@p=0\relax\else\hbox{--}\the\ncite@c\fi%
}

\def\refnum#1{\expandafter\csname\@citelabel{#1}\endcsname}

\catcode`@12
\catcode`@11
\def\biblabelcontents#1{{\bblrm$^{#1}$}}
%
\newcount\ftcount\ftcount1%
\def\ft#1{%
\ncite{ft:\the\ftcount}%
\expandafter\xdef\csname ftt:\the\ftcount\endcsname{#1\unskip}
\advance\ftcount1%
}
%
\def\bibhookdar#1#2{\expandafter\xdef\csname bibhookdar:#1\endcsname{#2}}
\def\bibhookdarexp#1{\ifx\expandafter\csname bibhookdar:#1\endcsname%
\undefined\relax\else\expandafter\csname bibhookdar:#1\endcsname\fi}

\catcode`@12
\twelvepoint
\def\mbox#1{\leavemode\hbox{#1}}	

\def\square{\hbox{
\vbox to.6em{%
\vfil%
\vbox{\boxx{\hbox{\vbox to0.3em{\hbox to0.3em{}}}}}%
\vfil%
}}}
\long\def\gencap#1#2{{\tenpoint%
\narrower\narrower\noindent#1\par}}
\long\def\figcap#1{\gencap{\expandafter\noexpand#1}{dummy}}
\long\def\tabcap#1{\gencap{\expandafter\noexpand#1}{dummy}}
\def\Rtop{R_{\scriptscriptstyle\rm top}}
\def\Rbot{R_{\scriptscriptstyle\rm bot}}
\clubpenalty9990
\widowpenalty9000
\def\doprint#1{\midinsert#1\endinsert}

\centerline{\bf Diagnosis and Location of Pinhole Defects in Tunnel Junctions}
\centerline{\bf using only Electrical Measurements}
\bigskip
\centerline{Zhongsheng Zhang and David A. Rabson}
\centerline{\it Department of Physics, PHY 114, University of South Florida,
Tampa, FL 33620, USA}
\smallskip
\bigskip

\centerline{\bf Abstract}
{\narrower\narrower%
In the development of the first generation of
sensors and memory chips based on spin-dependent tunneling
through a thin trilayer, it has become clear that
pinhole defects can have a deleterious
effect on magnetoresistance.  However, current diagnostic
protocols based on Andreev reflection and the temperature dependence of
junction resistance may not be
suitable for production quality control.
We show that the current
density in a tunnel junction in the cross-strip geometry
becomes very inhomogeneous in the presence of a single pinhole,
yielding a four-terminal resistance that depends on the
location of the pinhole in the junction.  Taking advantage
of this position dependence, we propose a simple protocol
of four four-terminal measurements.  Solving an inverse problem,
we can diagnose the presence of a pinhole and estimate its
position and resistance.
\par}

\bigskip
\leftline{\tt PACS '03: 73.40.Rw, 73.40.Jn, 85.75.Dd, 85.75.Mm}

\bigskip\bigskip\bigskip

Semiconductor manufacturers are currently developing
magnetic-random-access-memory elements \ncite{Parkin99,Daughton,Reohr02}
based on magnetic
tunnel junctions; such junctions separate two ferromagnetic
metallic leads by a thin insulting layer,
\ncite{Julliere75,Miyazaki95,Moodera99}
often
made by oxidizing a film of Al or other suitable metal.
Both because of the thinness of the insulating layer and because
of the possibility of inadequate oxidation, ``pinhole''
defects---direct metal-metal shorts through the nominal
insulator---have attracted significant 
attention.\ncite{Oepts98,Sousa99,Tatara99,Oepts99b,Jonsson00,Garcia00,Rabson01}
A single pinhole can also be generated in a previously pinhole-free
junction carefully through a voltage 
ramp\ncite{Oepts99,Shimazawa,Rao,Schmalhorst} 
or, by implication, inadvertently.
Generally, the parasitic current through pinholes detracts
from a junction's magnetoresistance,\ft{%
Reference \refnum{Garcia00} provides a theoretical mechanism
by which pinholes would {\it not\/} detract from magnetoresistance,
although several experimental investigations ({\it e.g.},
[\refnum{Sousa99},\refnum{Oepts99},\refnum{Shimazawa},\refnum{Rao}]) {\it do\/} see
a large effect.
} so methods for diagnosing
and locating such defects become important during the development
of practical devices.  Surprisingly, a fit of differential conductance
to the Simmons form \ncite{simmons} fails unambiguously to guarantee the 
absence of pinholes.\ncite{Jonsson00,Akerman02}
Surer methods include the use of an integrated superconducting
electrode,\ncite{Akerman02} the temperature dependence of
device resistance,\ncite{Akerman02,Akerman01,Rudiger01} and surface 
decoration.\ncite{Oepts98,Oepts99}

These diagnostics may not be integrated easily into a development
or manufacturing process.  We therefore propose a very simple
test that not only determines the presence of a pinhole with high confidence
but also can typically locate the pinhole to within 7\% of
the junction area and estimate the pinhole resistance.
We propose four four-terminal measurements.  A discrete
three-dimensional resistor model allows us to compute the result of each
measurement for an assumed pinhole position; working backward from a set of
four experimental measurements to the pinhole position therefore
constitutes an inverse problem, to which we demonstrate a solution.
Since the method applies to non-magnetic as well
as to magnetic tunnel junctions, we shall generally treat the junctions
without reference to magnetic properties.

We will consider the common cross-strip geometry of figure 1,
consisting of a conducting top strip separated from a perpendicular
conducting bottom
strip by a thin insulator.  The device's four leads are
numbered as in the figure.  In a typical four-terminal resistance
measurement, as shown, current is injected in lead 1 (top strip)
and removed through lead 4 (bottom strip), while the voltage is
measured between leads 3 and 2.  Moodera and 
collaborators \ncite{Moodera1,Moodera2} have pointed out 
that, because a conducting strip is not an isopotential,
such a four-terminal measurement will give
misleading, sometimes even negative, absolute resistances when
the resistance of the insulating layer is comparable to or smaller than
those of the upper and lower metals.  We generalize their observation for
a junction shorted by a single pinhole.

\doprint{
\centerline{\epsfxsize8.5cm\epsfbox{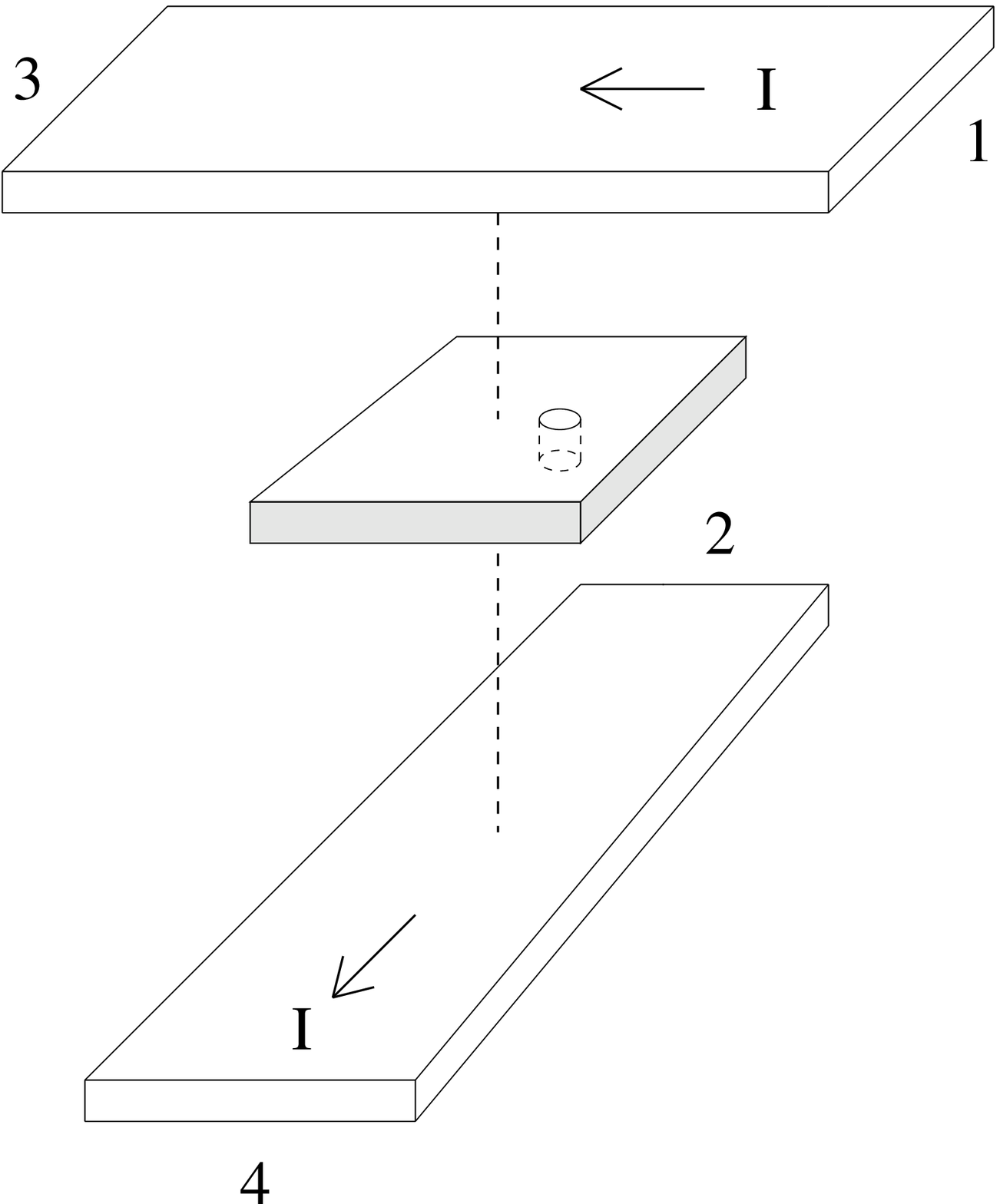}}
\figcap{Figure 1.  The standard cross-strip geometry (exploded view) consists of
a lower metallic strip over which is deposited an aluminum layer, which
is oxidized (shaded layer) before a top metallic strip is deposited.
To obtain the four-terminal resistance $R_1$, one injects current
through the leads labeled 1 and 4 while measuring the voltage between
leads 3 and 2.  A pinhole short through the insulating layer
will result in an unreliable four-terminal resistance that depends
on the position of the pinhole.
}
}

We model the junction
as a discretized three-layer resistor network.  Each of the
four long leads is assumed isopotential where it makes contact with
the square junction.  (Later, we address this assumption as a source
of uncertainty.)  The pinhole is modeled as a metallic (Al) inclusion
of diameter approximately $0.8\,$nm 
penetrating the insulating layer.
Classical conductivities are used for
conduction through metallic channels, including the pinhole, while
the very low conductance of tunneling is approximated from
Simmons's formula assuming a barrier height about 2$\,$eV and a thickness
around 1$\,$nm.
Kirchhoff's law yields a set of linear
equations for the potentials at all the nodes of the resistor
network in terms of fixed current
$I$ through two leads.  We solve the linear equations numerically
in order to calculate the voltage difference $V$ between the
two measurement leads. 

Denote pinhole resistance by $R_p$ and the intrinsic tunnel ``resistance,''
determined in the Simmons model by the barrier height, area, and thickness,
by $R_t$.
Figure 2 plots the nominal four-terminal resistance ($R_1$) against $R_t$
for $R_p$ fixed at $50\,\Omega$ (horizontal line); the top
and bottom layers are given
two-dimensional resistivities of
$\Rtop=10\,\Omega/\square$ and
$\Rbot=20\,\Omega/\square$.
These values appear representative of recent
experiments.\ncite{Moodera1,Moodera2}
The pinhole for this figure was placed in the bottom-left quadrant but not
very near the corner.
As noted,
the nominal four-terminal resistance takes anomalously small, even negative,
values when $R_t\ll \Rtop,\,
\Rbot$.  The four-terminal measurement is
independent of $R_t$ when the latter is large compared to the resistances
of the leads, since then most current flows through the pinhole.  One thus
measures a four-terminal resistance
$$
R_1(\bold r) = R_p + R_{1f}(\bold r)\rlap{\quad,}
\eq(R1)
$$
where $R_{1f}$ is a function of the pinhole's position, $\bold r$, for fixed 
$\Rtop$ and $\Rbot$.  In a junction of sufficiently small area,
$\Rtop,\,\Rbot,\,R_p\ll R_t$,
leaving the positional term,
$R_{1f}(\bold r)$, independent of $R_p$.
The inset to figure 2 shows voltage contours on the top layer; the pinhole
is evident.

\doprint{
\centerline{\epsfxsize8.5cm\epsfbox{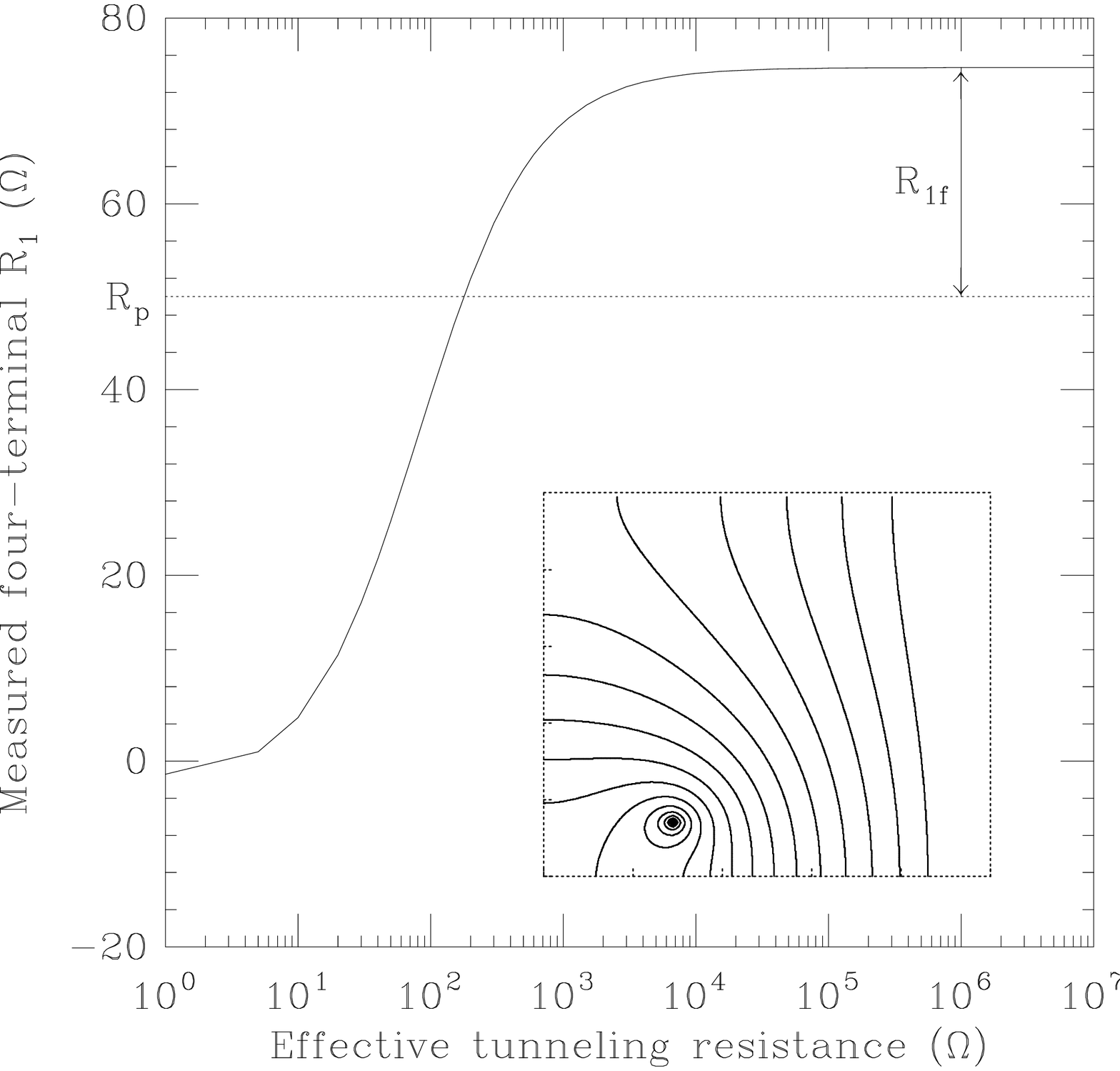}}
\figcap{Figure 2.  The four-terminal resistance of a junction with
a pinhole is plotted as a function of the effective resistance of the
tunneling layer.  For this illustration, the pinhole resistance
$R_p=50\Omega$, while the top- and bottom-layer resistivities
take the values $10\,\Omega/\square$ and $20\,\Omega/\square$.
When the tunneling resistance is sufficiently large,
most of the current flows through the pinhole, and the four-terminal
resistance is the sum of the pinhole resistance $R_p$ and a position-dependent
piece $R_{1f}(\bold r)$, as in \(R1).  The inset shows constant-voltage
contours on the top layer.
}
}

We now iterate over all possible pinhole positions, $\bold r$.
Figure 3 illustrates, for the same parameters as figure 2,
the position dependence of $R_1(\bold r)$.  If
we knew the value of $R_p$, this figure would localize the
pinhole to the curve of constant $R_1$ corresponding to the
measured value.

\doprint{
\centerline{\epsfxsize8.5cm\epsfbox{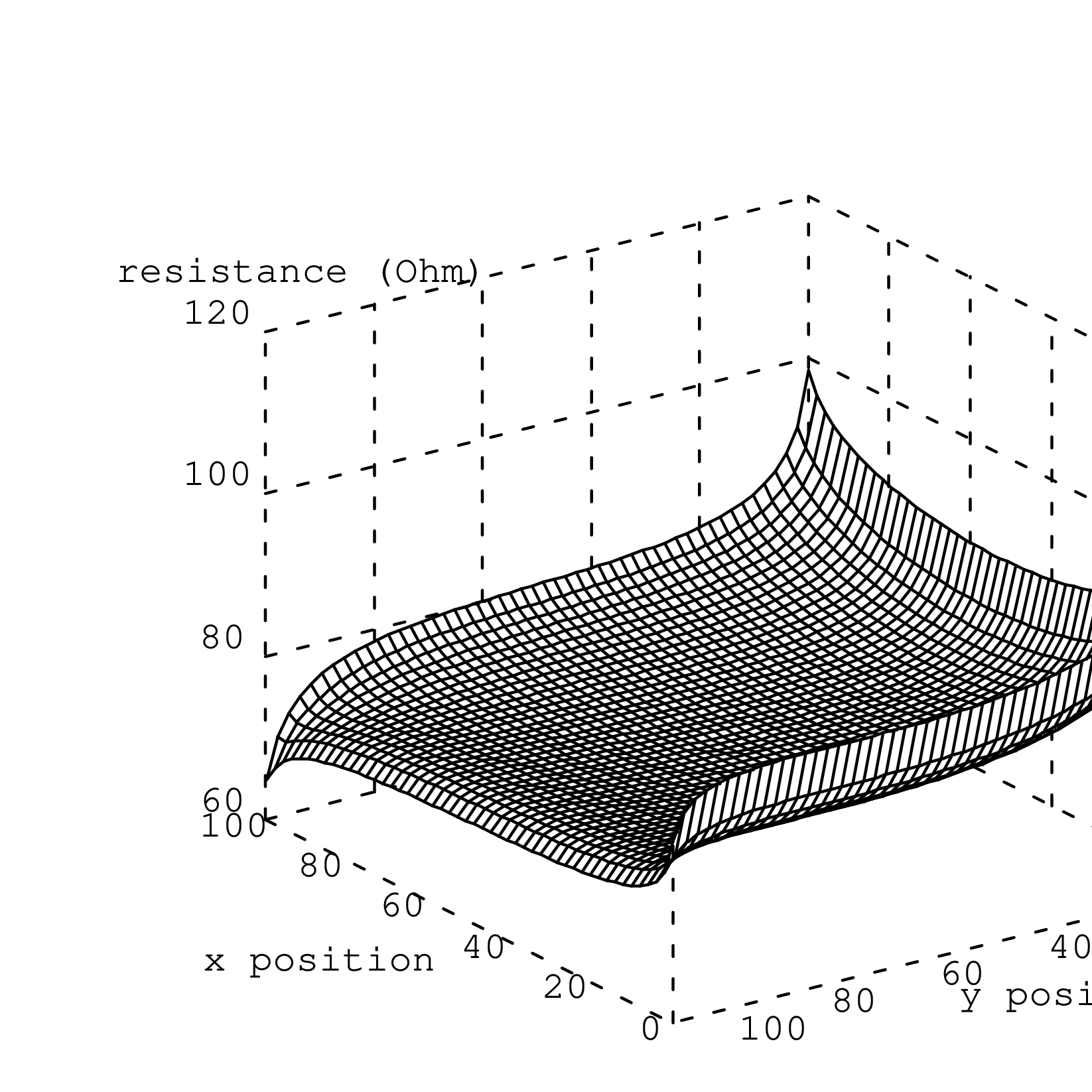}}
\figcap{Figure 3.  The predicted four-terminal resistance $R_1$ is plotted
against pinhole position.  By rotating the roles played by the four leads,
one can localize the pinhole over most of the junction area.
}
}

In fact, we can find both $R_p$ and $\bold r$ simply by rotating
the roles played by the four electrodes.  Thus we measure $R_2$
by forcing current through leads 1 and 2, measuring voltage across
leads 3 and 4.  Repeating the rotation, we get four four-terminal
measurements, as in table I.  In the absence of a pinhole,
these four
measurements would be equal.

\newbox\foobox\setbox\foobox\hbox{($R_1\,\dots\,R_4$)}%

\midinsert
{
\tabskip0pt\offinterlineskip
\def\tablerule{\noalign{\hrule}}
\def\strutt{\vtop to1.6\dp\strutbox{\null}\vbox to1.4\ht\strutbox{\null}}
$$
\vcenter{
\halign{\strutt#\hfil\ \ \vrule\ \ &#\hfil\ \ \vrule\ \ &#\hfil\ \cr
&\hfill $I$\hfill&\hfill $V$\hfill\cr
\tablerule
$R_1$ & $1,\,4$ & $3,\,2$\cr
$R_2$ & $2,\,1$ & $4,\,3$\cr
$R_3$ & $3,\,2$ & $1,\,4$\cr
$R_4$ & $4,\,3$ & $2,\,1$\cr
}
}
$$
\tabcap{Table I.  Four four-terminal
resistances \unhcopy\foobox\ are defined from the
four ways of injecting current ($I$) through one top and one bottom lead while
measuring the voltage ($V$) between the remaining two leads.}
}
\bigskip
\bigskip
\bigskip
\endinsert

From \(R1), differences of the resistances in table I will not
depend on the pinhole resistance $R_p$; it is convenient\ft{%
The square junction of figure 1 with no pinhole is invariant under
the point group $\roman D_{\rm 2d}$ ($=\bar42m$), which has a
four-dimensional representation on the original measurements
$R_i$, $i=1...4$.  This representation reduces to a one-dimensional
non-trivial representation under which $R_a$ is invariant, a
two-dimensional representation mixing $R_b$ and $R_c$, and a
trivial one-dimensional representation $R_d=R_1+R_2+R_3+R_4$,
only the last of which depends on $R_p$.  Of the other three,
only $R_a$ has two nodal lines, meaning that it will generally
be small compared to $R_b$ and $R_c$ except in the corners (see figure 4)}
to take as
a maximal independent set
$$
\eqalign{
R_a &~=~ R_1 - R_2 + R_3 - R_4\cr
R_b &~=~ R_1 + R_2 - R_3 - R_4\cr
R_c &~=~ R_1 - R_2 - R_3 + R_4\rlap{\quad.}\cr
}
\eq(Rabc)
$$

Figure 4 displays contour maps of each of the resistances from \(Rabc)
against position $\bold r$ of the pinhole,
with the contour corresponding to the (simulated) measurement
outlined.
If we may assume perfect knowledge of the geometry, the three outlined
contours will intersect at a point, thus locating the pinhole.

Measurements on an actual tunnel junction will suffer errors
and uncertainties from several sources, such as surface roughness and
possible non-percolating conducting inclusions.   The classical
model itself breaks down for junctions small compared to the relevant
mean free path.  We model the
uncertainties in two ways, chosen to be representative rather than
microscopically realistic.  We then propagate the uncertainties to
the calculation of resistances as a function of pinhole position.

First, we consider a break of up to 10\% of the junction
width at the point where one lead joins the junction, as though
there were a bad solder joint.  Of course, there is no solder---the
strip is continuous---but we use this 10\% void as a proxy for
uncertainty in the actual junction geometry.  A full calculation
would find equipotentials bending into the leads where the current
is injected rather than coming in straight; inhomogeneity in the leads
would further complicate the picture.
The 10\% void is varied over positions at
this junction edge, leading to an error in the four-terminal
measurement (compared to the result without the break).

Second, we posit a single atomic-scale terrace in the insulating layer;
this too is allowed to migrate over the junction, leading to an error.
This source of error stands in as well for possible tunneling ``hot spots''
due to roughness,\ncite{DaCosta98,Ando99,Buchanan02,Dorneles03}
so long as the tunnel current remains small
compared to the pinhole current.
The two sources of uncertainty are combined to give estimated error bars
in the resistances $R_a$, $R_b$, and $R_c$.  These error bars can
be pictured as broadening the outlined contours of figure 4,
and the single point of overlap becomes instead a region in which we
expect to find the pinhole.

\doprint{
\centerline{%
\kern-2.5em
\hfil\hfil%
\epsfxsize.31\hsize\epsfbox{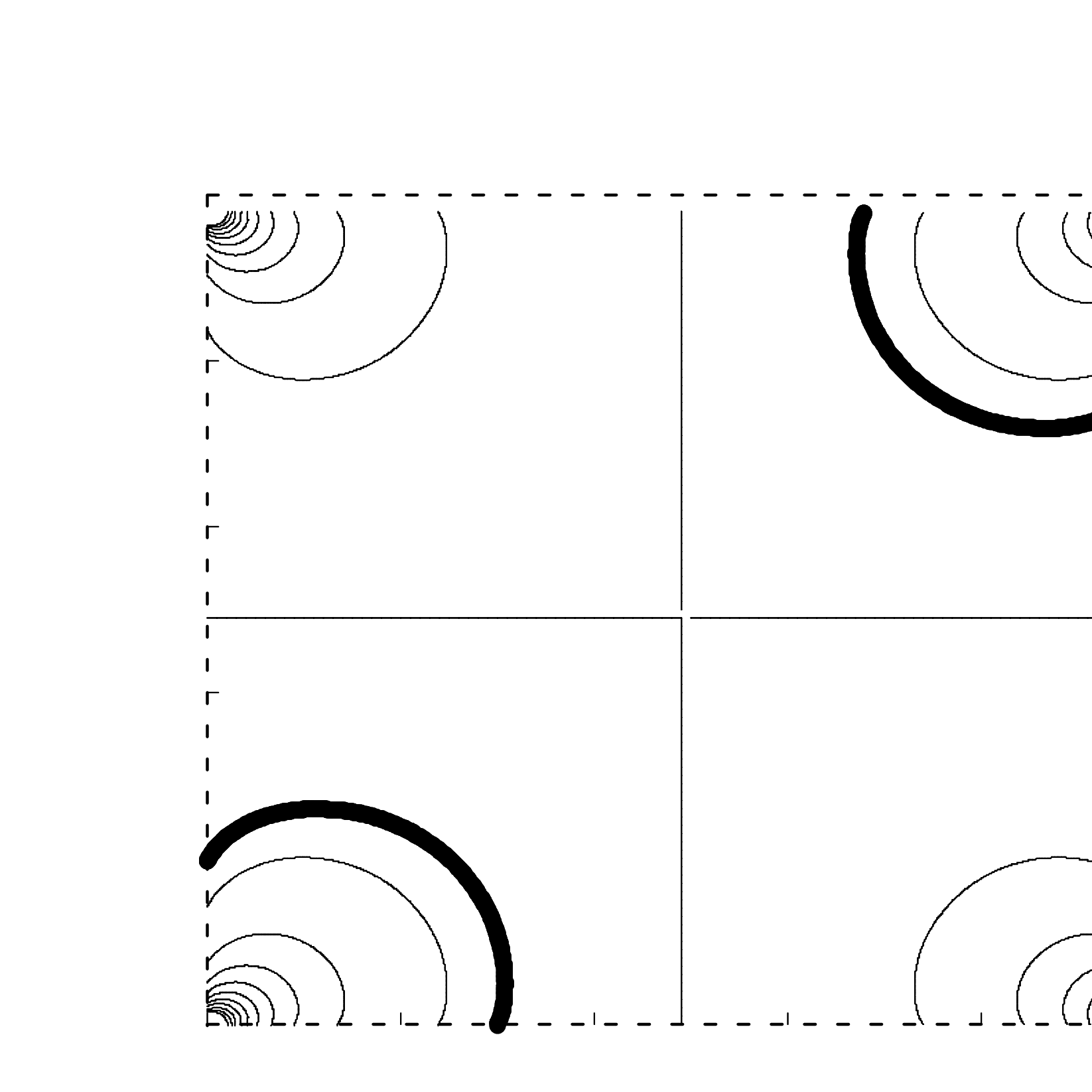}%
\hfil%
\epsfxsize.31\hsize\epsfbox{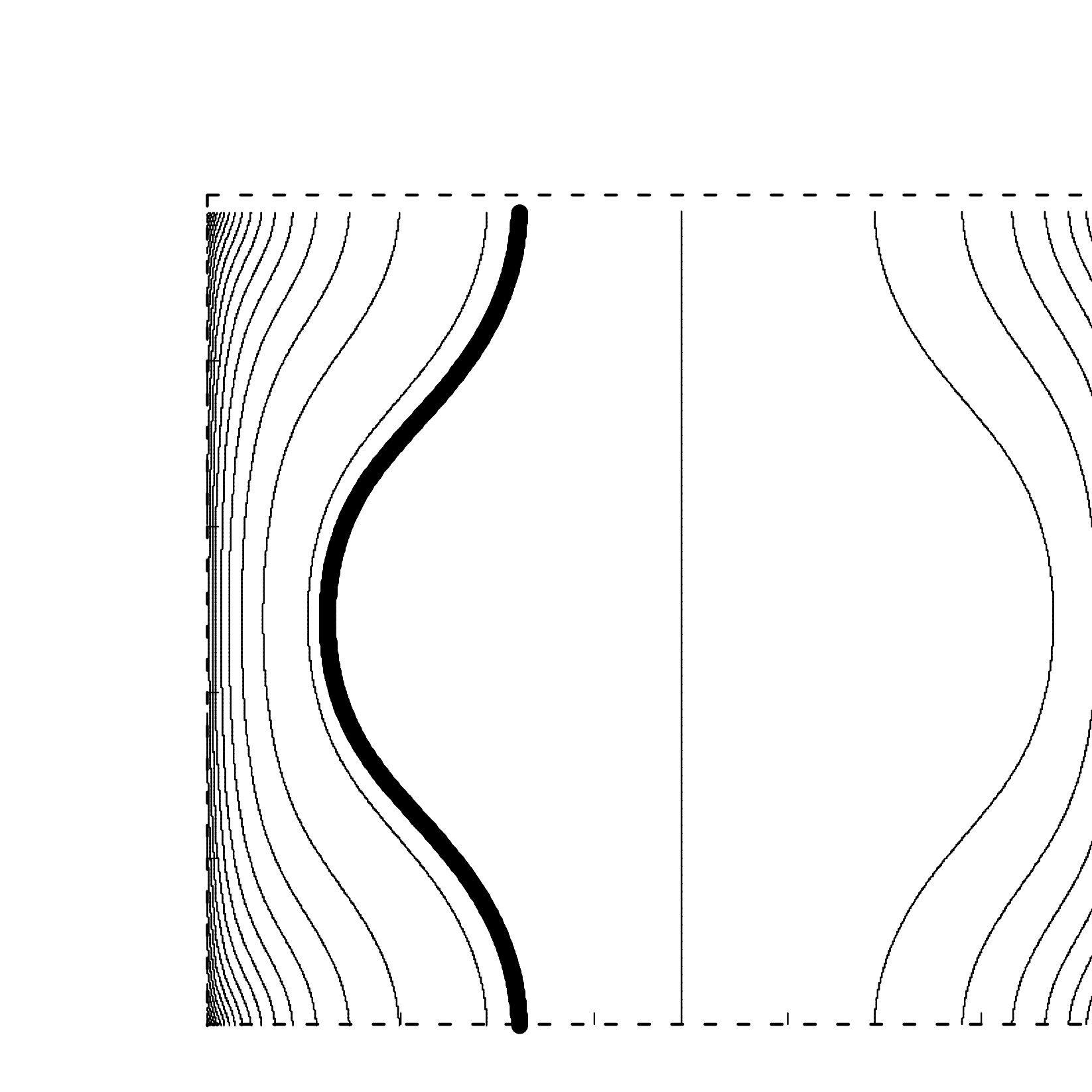}%
\hfil%
\epsfxsize.31\hsize\epsfbox{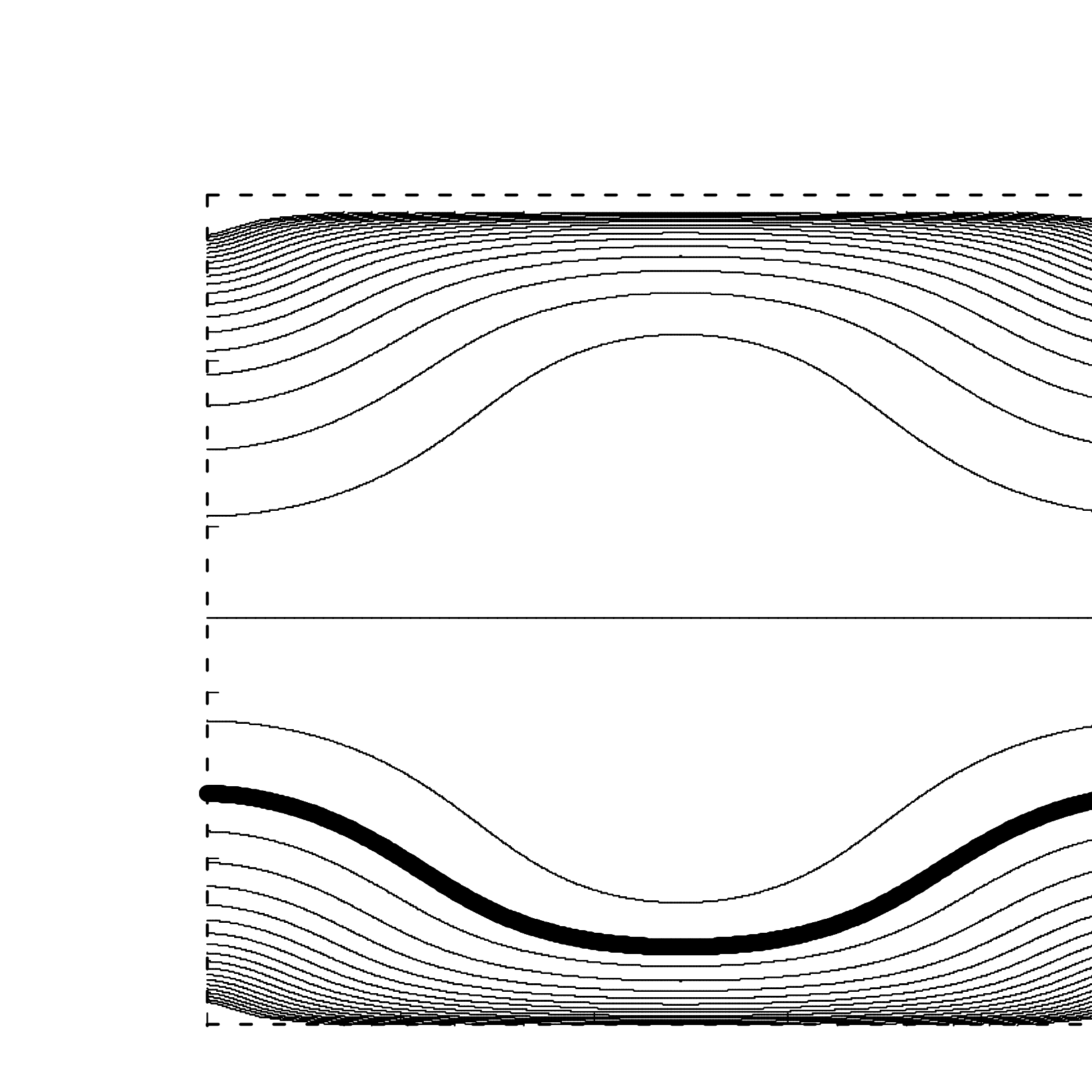}%
\hfil\hfil%
}
\figcap{Figure 4.  Resistance contours for $R_a$, $R_b$, and $R_c$
from equation \(Rabc).  The resistance that would be measured for the example
of figure 2
is emphasized in each.  The pinhole is located
at the intersection of the three emphasized contours.
Because of the two (perpendicular) node lines in
$R_a$, the values are much smaller:
the contour scale on $R_a$ is smaller by a factor of 20 than
the scales for the other two resistances.
}
}

For the sample geometry we have been considering, these sources of error
yield uncertainties in each of the four-terminal measurements of 0.5\%. 
If all four measurements $R_i$ (equation \(R1))
agree within this percentage,\ft{%
In the combinations \(Rabc), the errors are
correlated rather than independent, so we apply an error $\sigma=0.5\%$ instead
of twice this value}
we conclude with high confidence
that the junction does not harbor a pinhole.  (We discuss below
the probability of a false negative result.)

We have discussed the calculation of $R_i$ given a known
pinhole, but in fact we wish to determine the position of a pinhole
given only the measurements $R_i$.
We solve this inverse problem by brute-force computation of four-terminal
resistances for all possible positions of a single pinhole on a
two-dimensional grid.   Confidence regions, as in figure 5, are
determined by $\chi^2$ minimization and verified with Monte-Carlo
sampling.\ncite{NUMRECIPES}

\doprint{
\centerline{\epsfxsize8.5cm\epsfbox{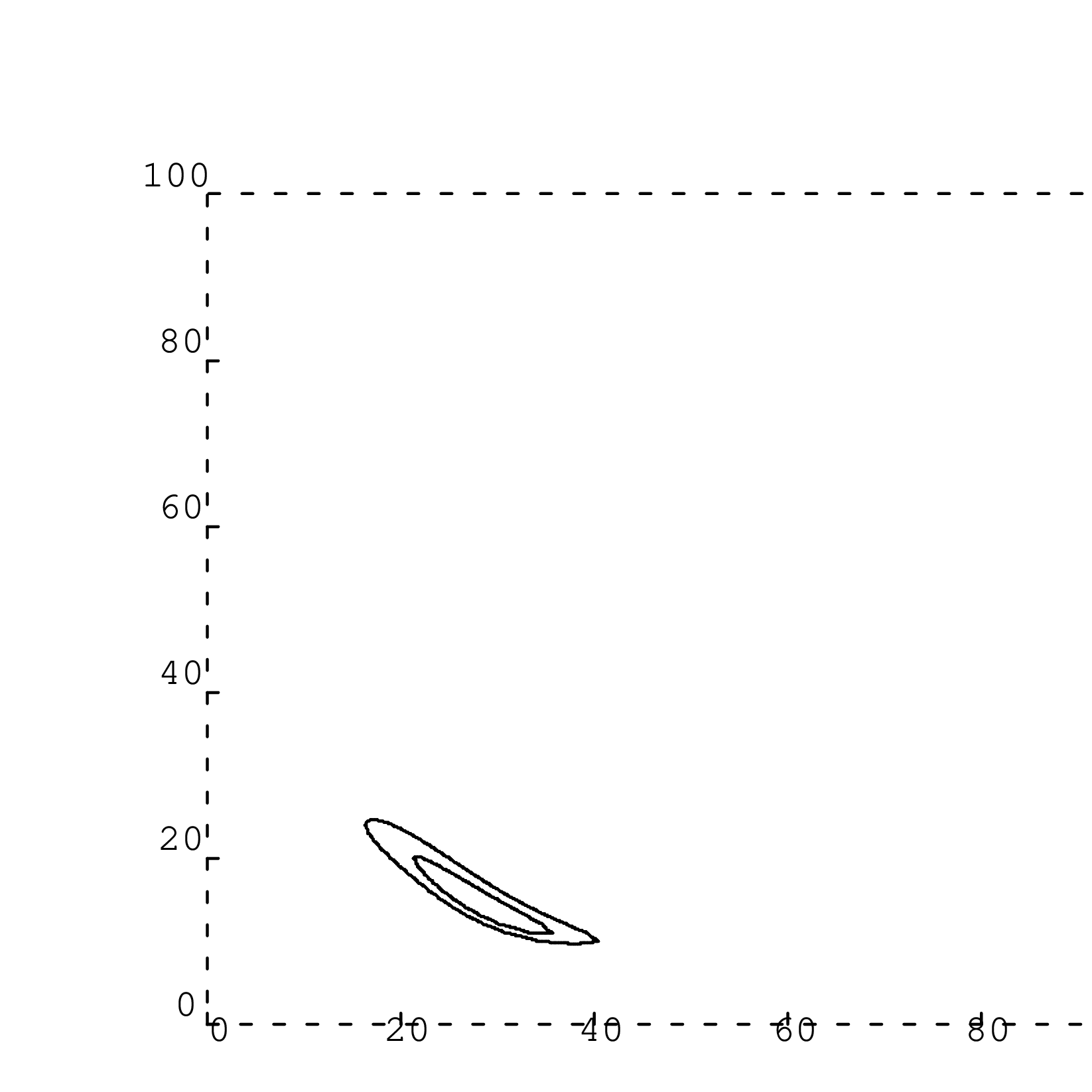}}
\figcap{Figure 5.  The pinhole of the example localized.  The two contours
bound 68\% and 95\% confidence regions as calculated with the $\chi^2$
method.  (Monte-Carlo gives similar pictures.)
}
}

To extract a confidence region from the $\chi^2$ method, we compare
experimental measurements $R_i^{(0)}$ to calculated resistances
$R_i(\bold r)$ as functions of pinhole position.  
The objective
function $\chi^2$ is
the sum of square errors between calculated and
measures resistances, normalized by the estimated variances $\sigma_i^2$:
$$
\chi^2(\bold r) ~=~
\sum_{i=a,b,c} \left(\left[R_i^{\vphantom{(0)}}(\bold r)
- R_i^{(0)}\right]/\sigma_i\right)^2
\rlap{\quad.}
\eq(chi2)
$$
We apply the usual chi-square distribution with two degrees of freedom
(corresponding to $\bold r$) to convert $\chi^2$
contours into confidence regions.

To verify these results with Monte-Carlo sampling, we
start with an assumed pinhole
position and resistance and calculate the four resistances $R_i^{(0)}$.
We then generate a large number of ``experimental'' data sets
$R_i^{(j)}$ by adding Gaussian-distributed noise according to the
standard deviations worked out previously.  For each data set,
we find the pinhole position $\bold r$ (on the discrete grid) that
minimizes $\chi^2$.  With a large number of sets $R_i^{(j)}$, we
accumulate the number of times each grid position minimizes the objective
function.  In the 68\% confidence region, we include first the grid
point with the largest count, then that with the second-largest count
{\it etc.}\ until we have included 68\% of the data sets.  The results
agree closely with those from the $\chi^2$ analysis.
The procedure could be refined by increasing the
number of grid points in the vicinity of the most probable pinhole position
until the resolved size fell below the dimensions of the 68\% (1-sigma)
confidence bound.

We have repeated this calculation for pinhole positions covering the
junction area.  As the four-fold symmetry makes evident, a pinhole
in the very center of the junction would be undetectable through this
method, since all four $R_i$ (equation \(R1)) would be equal.
Allowing for estimated
errors in the four measurements, we find a region covering about 13\%
of the area around the center in which a pinhole cannot
be distinguished from the absence of a pinhole:
a pinhole in this central region still yields measurements
$R_a,\,R_b,\,R_c$
all equal to zero within an uncertainty $2\sigma$.
Confidence regions for pinholes near the central region
are larger than for those closer to the edges and corners.
Considering all possible actual pinhole positions (including the center),
we find that, on
average, the 68\% confidence region comprises 3.5\% of
the junction area, while the 95\% confidence region comprises
7\%.
Having located the pinhole, we can solve for $R_p$
in \(R1) or, to take advantage of averaging, in $R_d=R_1+R_2+R_3+R_4$.

We have proposed a simple technique for diagnosing, with high
confidence, the presence or absence of a single pinhole in a tunnel
junction and furthermore for determining its position. 
We hope soon for an experimental test.
Pinhole-free junctions may be prepared (and verified using the
temperature dependence of resistance \ncite{Rudiger01,Akerman02} or absence of 
Andreev reflection \ncite{Akerman02}) and subjected to the protocol outlined here.
The four four-terminal measurements should be equal to within a
small uncertainty.  To the extent that this uncertainty, derived
from a number of junctions, differs from our rough estimate, it will
serve to recalibrate our estimated error bars.  Then, applying a slow
voltage ramp, a single pinhole can be created in each junction.  The
protocol should determine the pinhole's presence.  If possible, it
will be interesting to check actual pinhole location with a
decoration method \ncite{Oepts98,Oepts99} or by scanning-probe
microscopy.\ncite{DaCosta98,Wulfhekel}

Work now in preparation investigates the effects of Ohmic heating
and thermal transport on the differential conductance of a junction
incorporating both a pinhole and tunneling channels.  Further work
will replace the classical approximations made so far with quantum-mechanical
calculations.\ncite{Hershfield89}
\bibhookdar{Hershfield89}{As an example of a quantum-many-body investigation
of a four-terminal measurement, consider }

\bigskip
\centerline{\bf Acknowledgements}\nobreak
We thank Johan \AA kerman for useful comments.
DAR is a Cottrell Scholar of Research Corporation, which has supported
this work.

\bigskip\bigskip
\centerline{\bf References}\nobreak
\bibliography{pinhole}
\bibliographystyle{aip}

\vfil\eject
\end